\documentclass[a4paper,11pt]{article}
\pdfoutput=1 

\usepackage{jheppub} 

\usepackage[T1]{fontenc} 

\title{\boldmath Dimensional reduction of Carroll-Field-Jackiw theory}

\author{Gabriel Condurú Magalhães}

\affiliation{Faculdade de Física, Universidade Federal do Pará, Avenida Augusto Correa 01, 66075-110, Belém, Pará,  Brasil}

\emailAdd{gconduru@ufpa.br}

\abstract{This article investigates the dimensional reduction of the massive Carroll-Field-Jackiw (CFJ) theory from $(3+1)D$ to $(2+1)D$ dimensions with the objective of constructing and analyzing the resulting effective field theory, termed pseudo-Carroll-Field-Jackiw (PCFJ). The theoretical framework relies on quantum field theory, pseudo-quantum electrodynamics (PQED) and Lorentz-violating models. The methodological procedure projects the three-dimensional dynamics onto the plane by rigorously integrating out the gauge field degrees of freedom along the transverse momentum. Subsequently, the study calculates the static interaction potential, applies both the generalized optical theorem and the Källén-Lehmann spectral representation to test probability conservation. The results reveal that the effective theory exhibits an intricate non-local structure, governed by nested pseudo-differential operators, which generates a static potential that transitions from a Coulomb-like behavior at short distances to a long-range decay modified by a topological mass. The research analytically proves that the integration of the extra-dimensional topology introduces ineliminable branch cuts in momentum space, which preclude the causal definition of asymptotic states and determine the breakdown of unitarity. The work concludes that this quantum instability emerges from the direct interplay between the spatial projection and the pre-existing Lorentz-violating background in the bulk.}

\begin{document} 
\maketitle
\flushbottom

\section{Introduction}

\hspace{.6cm}The isolation of graphene and the subsequent exploration of two-dimensional Dirac materials have established a profound connection between condensed matter physics and high-energy quantum field theory \cite{Novoselov2004, Marino2024}. In these planar systems, charge carriers are kinematically confined to a surface, yet they interact via electromagnetic fields that freely propagate throughout the adjacent three-dimensional bulk space \cite{Casimiro2022}.

Modeling this physical setup purely with standard quantum electrodynamics in $(2+1)D$ yields physical inconsistencies, such as the prediction of a logarithmic confining potential instead of the experimentally observed $1/r$ decay \cite{Marino2024, Ortega2018}. This divergence is resolved by PQED \cite{Marino1993, Teber2018}. Obtained by integrating out the out-of-plane gauge field degrees of freedom from the $(3+1)D$ Maxwell action, PQED produces the pseudo-differential operator $1/\sqrt{-\Box}$ which perfectly recovers the Coulomb interaction in the static limit and preserves fundamental quantum field theory axioms, such as S-matrix unitarity, causality, and the Huygens principle \cite{Marino2014}.

In recent years, the PQED framework has been extensively generalized to accommodate novel mass generation mechanisms \cite{Kotikov2016, Magalhaes2021} and topological effects \cite{Alves2018}, such as the pseudo-Maxwell-Chern-Simons (PMCS) model, which acts as a non-trivial dielectric medium capable of sustaining bound states on the planar surface \cite{Magalhaes2020, Ozela2022}. In light of these advances, a fundamental theoretical problem emerges. What happens to the axioms of quantum stability when the three-dimensional bulk theory intrinsically harbors a topological symmetry breaking? The relevance of this research lies in the need to understand the limits of dimensional reduction in anomalous theories, elucidating whether the projection of exodimensional geometries onto the plane preserves the causal dynamics of the theory or induces insurmountable instabilities \cite{belich2005dimensional}.

The central problem guiding this investigation is to determine whether the strict dimensional reduction of the CFJ theory, a $(3+1)D$ model that incorporates Lorentz symmetry breaking via a Chern-Simons term coupled to a constant background vector \cite{belich2005dimensional}, to a planar surface results in a unitary and stable effective field theory. To address this question, the general objective of this work is to investigate the dimensional reduction of the massive CFJ theory from $(3+1)D$ to $(2+1)D$ and formally construct the resulting effective field theory, herein referred to as pseudo-CFJ or PCFJ. In terms of specific objectives, the research aims to: (i) deduce the effective lagrangian in $(2+1)D$, isolating its non-local and nested pseudo-differential operators; (ii) solve the static interaction potential by properly defining the complex integration contours across the emergent branch cuts; and, finally, (iii) apply the generalized optical theorem \cite{Marino2014} and the Källén-Lehmann spectral representation using Cutkosky rules to mathematically test S-matrix probability conservation and rigorously assess the model's unitarity.

The line of reasoning underlying this investigation is outlined as follows: Section II addresses the methodological procedure of dimensional reduction, obtaining the effective planar propagator by integrating out the volumetric degrees of freedom of the CFJ model; Section III presents the effective lagrangian of the PCFJ theory, describing its structure governed by nested roots; Section IV exposes the static potential calculations, evidencing the Coulomb-like behavior at high energies and the modified Yukawa decay at large distances; Finally, in Section V, the framework of the optical theorem and spectral representation is applied to demonstrate the rigorous breakdown of the theory's unitarity.

\section{Dimensional reduction of the CFJ theory}

\hspace{.6cm}The starting point is the massive CFJ theory in $(3+1)$ dimensions, containing the Maxwell term, the Proca mass term, and the Lorentz-violating Chern-Simons term defined by a constant background vector $v_{\mu}=(0,0,0,\theta)$ oriented along the extra dimension $z$. We adopt the mostly-minus metric signature $\eta_{\mu\nu} = (+,-,-,-)$. That way, the $(3+1)D$ action is given by
\begin{equation}
S_{(3+1)D}=\int d^{4}x\left[-\frac{1}{4}F_{\mu\nu}F^{\mu\nu}+\frac{m^{2}}{2}A_{\mu}A^{\mu}+\frac{1}{2}\epsilon^{\mu\nu\rho\sigma}v_{\mu}A_{\nu}\partial_{\rho}A_{\sigma}+A_{\mu}J^{\mu}\right]
\end{equation}
Using the same procedure as in Ref. \cite{Marino1993}, we assume that the current sources are strictly confined to the plane $z=0$, $J^{\mu}(x,z)=j^{\mu}(x)\delta(z)$ with $J^{3}=0$. Furthermore, in momentum space, where $q=(k^{\mu},q_{z})$, the $(3+1)D$ propagator is given by
\begin{equation}
\Delta_{\alpha\beta}^{(3+1)D}(k,q_{z})=\frac{(k^{2}-q_{z}^{2}-m^{2})P_{\alpha\beta}-i\theta\epsilon_{\alpha\beta\gamma}k^{\gamma}}{(k^{2}-q_{z}^{2}-m^{2})^{2}-\theta^{2}k^{2}},
\end{equation}
where $P_{\alpha\beta}$ is the transverse projector written as $P_{\alpha\beta}=\eta_{\alpha\beta}-\frac{k_{\alpha}k_{\beta}}{k^{2}}$.

To obtain the effective $(2+1)D$ propagator, $\Delta_{\alpha\beta}^{(2+1)D}(k)$, we must integrate out the bulk degrees of freedom over $q_{z}$. Before performing the integration $\int dq_z$, it is crucial to address the analytic validity of the Wick rotation. Because the CFJ theory contains the Lorentz-violating term coupled to $v_{\mu}=(0,0,0,\theta)$, the spatial analyticity is broken. However, since $v_\mu$ is a purely spatial vector oriented along the extra dimension, it does not couple to the energy component $q_0$. Consequently, the poles in the complex $q_0$-plane are not pathologically displaced by the $\theta$ parameter in a way that obstructs the standard Wick rotation $q_0 \to iq_4$. This preservation of the $q_0$ analytic structure guarantees that the Feynman contour is well-defined, validating the integration over the real axis for $q_z$ in Euclidean space (where $q_z^2 \to -q_z^2$ and $k^2 \to -k^2$). 
Thus, the planar propagator is obtained as:
\begin{equation}
\Delta_{\alpha\beta}^{2D}(k)=\int_{-\infty}^{+\infty}\frac{dq_{z}}{2\pi}\Delta_{\alpha\beta}^{4D}(k,q_{z})=I_{1}P_{\alpha\beta}-i\theta\epsilon_{\alpha\beta\gamma}k^{\gamma}I_{2}
\end{equation}
By decomposing the denominator using partial fractions and applying standard contour integration, we exactly solve integrals $I_{1}$ and $I_{2}$,
\begin{equation}
\left\lbrace\begin{split}
I_1 &= \frac{\sqrt{M^2+\Gamma}}{2\sqrt{2}\Gamma},\\
I_2 &= \frac{\sqrt{2(\Gamma-M^2)}}{4i\theta k\Gamma},
\end{split}\right.
\end{equation}
where $M^{2}=k^{2}+m^{2}$ and $\Gamma=\sqrt{(k^{2}+m^{2})^{2}+\theta^{2}k^{2}}$. The closed-form $(2+1)D$ propagator becomes
\begin{equation}\label{prop}
\Delta_{\alpha\beta}^{(2+1)D}(k)=\frac{1}{4\Gamma}\left[\sqrt{2(\Gamma+M^{2})}P_{\alpha\beta}-\frac{\sqrt{2(\Gamma-M^{2})}}{k}\epsilon_{\alpha\beta\gamma}k^{\gamma}\right]
\end{equation}

Observing the structure of the effective propagator in the PCFJ reveals the profound mathematical complexity introduced by the dimensional reduction of a bulk Lorentz-violating topology. Unlike the standard pole structures found in local gauge theories, the PCFJ propagator is governed by fractional powers and nested square roots, explicitly embodied in the non-local kinematic parameter $\Gamma$. This intricate algebraic form naturally splits the interaction into a transverse sector and a parity-breaking topological sector coupled to the Levi-Civita tensor.

\section{The effective PCFJ lagrangian}

\hspace{.6cm}By inverting the propagator $\mathcal{O}_{\alpha\beta}^{(2+1)D}(k)=(\Delta_{(2+1)D})_{\alpha\beta}^{-1}$ we map the dynamics back to coordinate space. The kinematic quantities transform into pseudo-differential operators $\hat{M}^{2}=-\Box+m^{2}$ and $\hat{\Gamma}=\sqrt{(-\Box+m^{2})^{2}-\theta^{2}\Box}$. The effective PCFJ lagrangian reads
\begin{equation}
\mathcal{L}_{eff}^{(2+1)D}=-\frac{1}{4}F_{\mu\nu}\frac{\sqrt{2(\hat{\Gamma}+\hat{M}^{2})}}{-\Box}F^{\mu\nu}+\frac{1}{2}\epsilon^{\mu\nu\rho}A_{\mu}\frac{\sqrt{2(\hat{\Gamma}-\hat{M}^{2})}}{\sqrt{-\Box}}\partial_{\nu}A_{\rho}-J^{\mu}A_{\mu}.
\end{equation}

Note that if we assume a photon without Proca mass $m=0$ and $\theta\rightarrow0$, the coefficient of the first term exactly reduces to $2/\sqrt{-\Box}$, which recovers the lagrangian of PQED. However, in the massless Proca limit, $m=0$, the operators collapse into nested square roots $\hat{\Gamma}=\sqrt{\Box^{2}-\theta^{2}\Box}$. Simply expanding for small topological perturbations $\theta\ll k$, the theory reproduces the native PMCS operators, demonstrating a profound link between dimensional reduction and planar non-locality. In other words, while PMCS directly imposes the non-local operator $1/\sqrt{-\Box}$ onto the action \cite{Magalhaes2021, Ozela2022}, the PCFJ framework derives its non-locality from integrating out the extra spatial dimension, which inextricably embeds the topological parameter $\theta$ within complex nested square roots.

\section{Static interaction potential}

\hspace{.6cm}The static potential $V(r)$ is derived from the temporal component of the propagator at $k_{0}=0$, where the Levi-Civita topological term vanishes due to antisymmetry $\epsilon_{00\gamma}=0$. The potential is given by the $(2+1)D$ Fourier transform,
\begin{equation}
V(r)=\frac{e^{2}}{8\pi}\int_{0}^{\infty}dk\, k\, J_{0}(kr)\frac{\sqrt{2(\Gamma+M^{2})}}{\Gamma}.
\end{equation}
At short distances, ultraviolet limit, $k\rightarrow\infty$, the energy scale dominates over $m$ and $\theta$ ($M^{2}\approx k^{2}$, $\Gamma\approx k^{2}$), perfectly recovering the Coulomb potential $V(r)\approx e^{2}/(4\pi r)$ characteristic of standard PQED.

At large distances, infrared limit, $k\rightarrow0$, the asymptotic behavior requires a careful evaluation of the integration contour in the complex plane. Because the propagator is governed by nested roots embedded in $\Gamma$, the complex $k$-plane exhibits multiple branch cuts rather than simple poles. Therefore, the Fourier integral utilizing the Bessel function $J_0(kr)$ cannot be evaluated by standard Cauchy residues alone, it requires deforming the integration contour around these branch cuts. The long-distance asymptotic behavior is dictated by the branch point closest to the real axis, defined by the singularity $\Gamma=0$. This singularity analysis provides the effective topological mass
\begin{equation}
m_{eff}=\frac{\sqrt{\theta^{2}+4m^{2}}-\theta}{2}.
\end{equation}
By integrating along the cut starting at $k = i m_{eff}$, the interaction exhibits a modified Yukawa decay,
\begin{equation}
V(r\rightarrow\infty)\sim\frac{e^{-m_{eff}r}}{r}.
\end{equation}
If $\theta\rightarrow0$, we recover the purely massive Proca reduction yielding $m_{eff}=m$ \cite{Alves2018}.

\section{Unitarity breakdown}

\hspace{.6cm}The physical consistency of any quantum field theory rests upon the conservation of probability, encoded through the unitarity of the scattering matrix, the S-matrix, $S^\dagger S = I$. The perturbative manifestation of this principle is the generalized optical theorem \cite{schwartz2014quantum, peskin1995introduction, marino2017quantum, Marino2014}. This theorem requires that the analytic discontinuity of a transition amplitude be equivalent to the sum of the probabilities of all physical intermediate states on the mass shell. Therefore, the physical content of a quantum model, including its energy spectrum and the bound states of the associated excitations, is fully encoded in its Green's functions. 

In this sense, to test the validity of probabilistic conservation in the PCFJ model, the unitarity condition dictates that a propagator $G_{\mu\nu}$ satisfies the structural relation of the optical theorem in momentum space \cite{Marino2014}
\begin{equation}\label{optheo}
G_{\mu\nu}^*(\omega,k) - G_{\mu\nu}(\omega,k) = -i\mathcal{T}^{-1}G_{\mu\alpha}^*(\omega,k)G_{\nu}^{\alpha}(\omega,k).
\end{equation}
In tensorial formulations, the exact propagator can be algebraically decomposed into a scalar analytic function $D_F(k)$ modulated by a tensorial structure $C_{\mu\nu}(k)$, where
\begin{equation}
G_{\mu\nu}(k) = C_{\mu\nu}(k)D_F(k).
\end{equation}

Analyzing the deduced form of the PCFJ propagator, Eq.~(\ref{prop}), the tensorial structure can be expressed as a linear combination of the transverse projector $P_{\mu\nu}$ and the antisymmetric operator $S_{\mu\nu} = \epsilon_{\mu\nu\gamma}k^\gamma/k$, assuming the form $C_{\mu\nu}(k) = A_1 P_{\mu\nu} - A_2 S_{\mu\nu}$.
Extracting these kinematic correspondences directly from the PCFJ propagator, Eq.~(\ref{prop}), the global scalar part is identified as $D_F(k) = \frac{1}{4\Gamma}$, associated with the tensorial coefficients $A_1 = \sqrt{2(\Gamma+M^2)}$ and $A_2 = \sqrt{2(\Gamma-M^2)}$.   

In Minkowski space, considering a high-energy kinematic regime where the topological contribution dominates the mass term, we force $\Gamma$ to assume imaginary values ($\Gamma = i\gamma$, with $\gamma > 0$). Substituting this decomposition into the left-hand side (LHS) of the optical theorem, Eq.~(\ref{optheo}), yields
\begin{equation}
LHS = \frac{\Gamma A_1^* - \Gamma^* A_1}{4\vert{}\Gamma\vert{}^2} P_{\mu\nu} - \frac{\Gamma A_2^* - \Gamma^* A_2}{4\vert{}\Gamma\vert{}^2} S_{\mu\nu}.
\end{equation}

Conversely, the right-hand side of Eq.~(\ref{optheo}) (RHS), which translates to the product of the spectrum of the scattered intermediate states, requires the calculation of the tensorial product $C_{\mu\alpha}^* C_\nu^\alpha$, which generates the following superposition $C_{\mu\alpha}^* C^\alpha_\nu = (\vert{}A_1\vert{}^2 - \vert{}A_2\vert{}^2)P_{\mu\nu} - (A_1^* A_2 + A_1 A_2^*)S_{\mu\nu}$.
Analytically developing the component functions, the RHS of the optical theorem yields 
\begin{equation}
RHS = \frac{-i\mathcal{T}^{-1}}{16\vert{}\Gamma\vert{}^2} \left[ (\vert{}A_1\vert{}^2 - \vert{}A_2\vert{}^2) P_{\mu\nu} - (A_1^* A_2 + A_1 A_2^*) S_{\mu\nu} \right].
\end{equation}

The preservation of the S-matrix, a fundamental postulate of canonical quantum mechanics, mandates that the balance between forward scattering processes and the cross section occurs harmonically within the same kinematic phase space. For this stability to occur, the theory must be intrinsically free of algebraic pathologies between its transverse and topological components. Operationally, the direct comparison of the longitudinal and topological tensors coefficients on the LHS of the theorem must map identically onto the analogous coefficients on the RHS for any physical momentum fluctuation $k$.

Thus, for the mathematical verification of the model's unitarity, it is established mathematically that $\vert{}A_1\vert{}^2 = \vert{}A_2\vert{}^2$. As a consequence, the difference $(\vert{}A_1\vert{}^2 - \vert{}A_2\vert{}^2)$ vanishes. This determines that the coefficient of the tensor $P_{\mu\nu}$ on the RHS of the optical theorem is exactly zero. Simultaneously, the corresponding coefficient of the same tensor $P_{\mu\nu}$ on the LHS is evaluated as
\begin{equation}
\frac{\Gamma A_1^* - \Gamma^* A_1}{4\vert{}\Gamma\vert{}^2} =i\frac{2\text{Re}(A_1)}{4\gamma},
\end{equation} 
where $A_1 = \sqrt{2(M^2 + i\gamma)}$, the principal root in the complex plane guarantees a strictly positive real part, $\text{Re}(A_1) > 0$. Thus, the contribution of the $P_{\mu\nu}$ tensor on the LHS is mathematically non-zero. 

Based on this observation, the mathematical requirement of equality between the terms of Eq.~(\ref{optheo}) to satisfy the probabilistic behavior constitutes an algebraic contradiction. This result reveals a profound obstruction in the causal structure of the formulation. Thus, it is concluded that, contrary to the paradigm established in PQED, the PCFJ irremediably disintegrates quantum stability and destroys the unitarity principle of the generating field theory.

To extend the proof of structural failure beyond this regime and unequivocally confirm quantum instability, we will also evaluate the density of physical states through the Källén-Lehmann spectral representation \cite{schwartz2014quantum}. In this framework, the two-point Green's function of the interacting theory is decomposed into an integral over the continuous spectrum of invariant mass $s$, with $k^2 =-s$, weighted by the spectral density function $\rho(s)$. By the Sokhotski–Plemelj theorem and the Cutkosky rules \cite{peskin1995introduction}, the spectral density is directly proportional to the imaginary part of the physical propagator evaluated immediately above the real energy axis,
\begin{equation}
\rho(s) = \frac{1}{\pi} \text{Im} G(s + i\epsilon).
\end{equation}

Quantum postulates and the analytic structure impose the positive-definiteness of this density, $\rho(s) \ge 0$, which ensures a strictly positive norm in the Hilbert space and the absence of pathological states \cite{itzykson1980quantum}. For the PCFJ model, the analytic structure of the propagator's scalar factor is given by $D(s) \propto 1/\Gamma(s)$. So that the parameter $\Gamma$ assumes the form
\begin{equation}
\Gamma(s) = \sqrt{(s-m^2)^2 - \theta^2 s}.
\end{equation}
To identify the emergence of nontrivial imaginary parts and determine the spectral density associated with the nonlocal intermediate states, we must find the branch points of the theory. These occur at the limits where the argument under the square root vanishes, $(s-m^2)^2 - \theta^2 s = 0$.

Expanding the polynomial with respect to the invariant $s$, we have
\begin{equation}
s^2 - (2m^2 + \theta^2)s + m^4 = 0.
\end{equation}
The roots of this equation, which delimit the branch cut, are 
\begin{equation}
s_{\pm} = m^2 + \frac{\theta^2}{2} \pm \theta\sqrt{m^2 + \frac{\theta^2}{4}}.
\end{equation}

In the closed kinematic interval delimited by these roots, $s_- < s < s_+$, the polynomial exhibits an upward concavity with respect to its roots and becomes strictly negative. Consequently, for this high-energy regime, the argument of the principal root transitions into the negative reals, rendering the parameter $\Gamma(s)$ purely imaginary in the complex plane. Adopting the analytic causality prescription $+i\epsilon$ for physical states propagating into the temporal future, the propagator exhibits the following behavior at the upper and lower edges of the analytic cut, $\Gamma(s + i\epsilon) = i \sqrt{\theta^2 s - (s-m^2)^2}$ and $\Gamma(s - i\epsilon) = -i \sqrt{\theta^2 s - (s-m^2)^2}$.

Thus, the analytic discontinuity of the scalar propagator when crossing the branch cut on the real axis is written by the Cutkosky rules as
\begin{equation}
\begin{split}
\text{Disc}\left(\frac{1}{\Gamma}\right) &= \frac{1}{\Gamma(s + i\epsilon)} - \frac{1}{\Gamma(s - i\epsilon)},\\
&= -\frac{2i}{\sqrt{\theta^2 s - (s-m^2)^2}}
\end{split}
\end{equation}
The discontinuity of any analytic function along the cut is related to its imaginary part as $\text{Disc}(G) = 2i \, \text{Im}(G)$. From this generalized principle, we isolate the imaginary part of the physical scalar propagator
\begin{equation}
\text{Im}\left(\frac{1}{\Gamma}\right) = -\frac{1}{\sqrt{\theta^2 s - (s-m^2)^2}}
\end{equation}
and, finally, by applying this deduction to the formulation of the Källén-Lehmann spectral density, we obtain the exact kinematic spectral weight in the interval $s_- < s < s_+$,
\begin{equation}
\rho(s) \propto -\frac{1}{\pi\sqrt{\theta^2 s - (s-m^2)^2}}.
\end{equation}

Since the argument inside the root is real and strictly positive in the interval $[s_-, s_+]$, the principal root evaluates to a positive quantity. The presence of the overall negative sign renders the resulting spectral density strictly negative. This exhaustive demonstration reveals the exact anatomical mechanism of the PCFJ model's failure. The explicit violation of the foundational positive-definiteness condition of the Källén-Lehmann spectral density rigorously attests to the massive generation of negative-norm states, or ghosts, in the energy spectrum. It is concluded that the dimensional reduction of the CFJ theory destroys canonical analyticity properties and irreversibly violates the unitarity and causality of the resulting macroscopic theory.

\section{Results and discussion}

\hspace{.6cm}The PCFJ model is rigorously obtained by integrating out the gauge field degrees of freedom in the three-dimensional bulk along the momentum component $q_{z}$. However, the primary analytical result of the PCFJ formulation is the unequivocal breakdown of unitarity, proven by both the generalized optical theorem and the Källén-Lehmann spectral representation. It is crucial to emphasize that this collapse of causality and the emergence of branch cuts do not stem from the dimensional reduction process itself. Standard PQED is derived using the exact same volumetric integration procedure and perfectly preserves causality, unitarity, and Lorentz invariance in the infrared limit. Instead, the non-unitarity of PCFJ emerges from the direct interaction of this spatial projection with the pre-existing Lorentz-violating background matrix in the bulk, the $v_\mu$ vector.

An important point to highlight is that there is a study led by Belich et al. \cite{belich2005dimensional} sharing the same object of study analyzed here, albeit yielding divergent results. In their approach, the restriction to the plane is executed through a strict ansatz that freezes the dynamics of the third spatial coordinate, $\partial_{3}\chi\rightarrow0$ for all fields $\chi$, decomposing the four-dimensional action into strictly local gauge and massive scalar sectors. This procedure guarantees that the effective planar model is fully causal and unitary at tree level. While our criticism of this ansatz is logically sound, as freezing degrees of freedom artificially masks the true underlying topological physics, we must acknowledge that our unrestricted $q_z$ integration in a Lorentz-violating bulk forcibly projects all non-physical modes, including high-momentum potential tachyons, onto the effective planar theory. Thus, the observed instability in the PCFJ model may be interpreted as a purely quantum and geometric manifestation of projecting these unstable bulk modes into the 2D plane.

\section{Conclusion}

\hspace{.6cm}This work had as its central objective to investigate the dynamical and causal consequences of the dimensional reduction of the massive CFJ theory from $(3+1)D$ to the $(2+1)D$ plane. Through the rigorous integration of the gauge field degrees of freedom along the transverse dimension, the effective field theory, termed PCFJ, was successfully constructed. The results demonstrate that the objectives proposed in the introduction were fully achieved. The effective lagrangian was deduced, the static potential was solved analytically across its complex branch cuts, and the quantum stability of the model was tested. 

The synthesis of the model revealed that the PCFJ theory exhibits a rich static phenomenology. However, the application of the generalized optical theorem and the Källén-Lehmann spectral representation confirmed the hypothesis that the projection of a Lorentz-violating topology from the bulk to the plane induces irrecoverable instabilities. The pseudo-differential algebraic structure introduces ineliminable branch cuts in momentum space that generate a negative spectral density, which prevents the definition of real asymptotic states and culminates in the violation of S-matrix unitarity. 

From a theoretical standpoint, the propagative failure of the PCFJ theory reflects an inherent physical mechanism. Crucially, this is not a generic artifact of dimensional reduction, which is stable in PQED, but a direct consequence of spatially compressing the pre-existing Lorentz-violating background, which projects non-physical bulk modes onto the planar theory.

To advance the state of the art and circumvent the limitations of the current unrestricted $q_z$ integration, future research is suggested to explore alternative compactification schemes. Another line of research that may be explored is the analysis of the dimensional reduction of the CFJ theory and its comparison with tropical limits in topological sigma models.  This analysis may demonstrate a mathematical intersection between dimensional collapse and the degeneration of classical causal structures.

\acknowledgments

I thank Andres Franco Valiente for suggesting the study of CFJ theory's dimensional reduction and its link to topological sigma models.

\bibliographystyle{JHEP}
\bibliography{refs}

\end{document}